# Image Restoration: A Comparative Analysis of Image De noising Using Different Spatial Filtering Techniques

**Onyedinma, E. G. and Onyenwe, I. E.**

*Department of Computer Science,NnamdiAzikiwe University, Awka. Anambra State, Nigeria*

**Abstract:** Acquired images for medical and other purposes can be affected by noise from both the equipment used in the capturing or the environment. This can have adverse effect on the information therein. Thus, the need to restore the image to its original state by removing the noise. To effectively remove such noise, pre knowledge of the type of noise model present is necessary. This work explores different noise removal filters by first introducing noise to an image and then applying different spatial domain filtering techniques to the image to get rid of the noise. Different evaluation techniques such as Peak to Signal Noise Ratio(PSNR) and Root Mean Square Error(RMSE) were adopted to determine how effective each filter is on a given image noise. Result showed that some filters are more effective on some noise models than others.

**Keywords:** Restoration, Degradation, Filter, Pixel. Blur.

**I. Introduction**

Image restoration is an objective process of improving an image for a predefined purpose. It tends to recover an image that has been degraded by using a priori knowledge of the degradation phenomenon. Degradation may be as a result of motion blur or noise. In degradation due to motion blur, it is possible to come up with a very good estimate of the actual blurring function and "undo" the blur to restore the original image. Conversely, where the image is corrupt by noise, the closest we may hope to achieve is to compensate for the degradation it caused. Thus, restoration techniques are oriented toward modeling the degradation and applying the inverse process in order to recover the original image [1]. This can be achieved by blurring or denoising. Blurring is making the image less clear or distinct by smoothing it.

In this paper, the model of the image degradation/restoration explored comprises:

- A degradation function $H$
- An additive/multiplicative noise: $\eta(x,y)$
- Input image: $f(x,y)$
- Degraded image: $g(x,y)$

Given the degraded image $g(x,y)$, some knowledge about the degradation function H, and some knowledge about the additive/multiplicative noise; the objective of restoration is to obtain an estimate $\hat{f}(x,y)$ of the original image. The more knowledge of H and η, the closer $\hat{f}(x,y)$ will be to $f(x,y)$.

The degraded image in the spatial domain (*if h is linear and position invariant process*) is given by:

$$g(x,y) = h(x,y) \star f(x,y) + \eta(x,y) \quad (1)$$

Where $h(x,y)$ is the spatial representation of the degradation function and $\star$ indicates convolution. By spatial domain, we mean the plane containing the pixels of an image.

Operations in the spatial domain are performed directly on the pixels of the image. Spatial domain is considered in this work because it is computationally efficient and requires less processing resources as compared with the Fourier transform.

**II. Noise Models**

Noise is present in all images captured by real world sensors[11]. Image noise is a random variation of brightness or color information in images, and is usually present in digital images during image acquisition, coding, transmission, and processing steps [5].These noises may result from noise sources present in the vicinity of the image capturing devices, faulty memory location, due to imperfection/inaccuracy in the image capturing devices like cameras, misaligned lenses, weak focal length, scattering and other adverse atmospheric conditions. Image noise can range from nearly invisible specks on a digital snapshot taken in good lighting to optical and radio astronomical images that are almost noise, from which a small amount of information can be extracted by complex processing [2]. Noise produces undesirable effects such as artifacts, unrealistic edges, unseen lines, corners, blurred objects and disturbs background scenes [3] that obscure the desired information. Consequently, the need for image de noising.

Image denoising is the first preprocessing step in image processing. To denoisean image, the image is processed using certain restoration techniques to remove induced noise which may creep into the image during acquisition, transmission or compression process. To understand the denoising process, it is essential to understand the noise function $\eta(x, y)$. Good knowledge of the noise function will determine the appropriate noise model to be deployed. More so, noise being by nature a random phenomenon, is modeled by a probability density which represents the intensity distribution of the noise. There are different noise models of which the noise removal techniques must be chosen according to the degree of image quality degradation as well as the type of noise present in the image [6]. The model could be additive or multiplicative. In the Additive Noise Model, an additive noise signal is added to the original signal to produce a corrupt noisy signal that follows the following rule:

$$g(x, y) = f(x, y) + \eta(x, y) \qquad (2)$$

where, $f(x, y)$ represents the original image intensity and $\eta(x,y)$ represents the noise supplied to produce the corrupt signal $g(x,y)$ at $(x,y)$ pixel position. Similarly, the Multiplicative Noise Model multiplies the original signal by the noise signal:

$$w(x, y) = f(x, y) * \eta(x, y) \qquad (3)$$

The following section discusses different noise models that can be applied in image processing.

**2.1 Gaussian noise**

Gaussian noise, also known as normal noise model is additive in nature [6] and follows Gaussian distribution. This implies that each pixel in the noisy image is the sum of the true pixel value and a random Gaussian distributed noise value. This noise is independent of intensity of the pixel value at each point. The PDF (Probability Density Function) of Gaussian random variable, z, is given by:

$$p(z) = \frac{1}{\sqrt{2\pi}\sigma} e^{-(z-\bar{z})^2/2\sigma^2} \qquad (4)$$

Where z represents intensity, $\bar{z}$ is the mean (average) value of z, and $\sigma$ is its standard deviation. The standard deviation squared, $\sigma^2$, is called the variance of z. The function is depicted in figure 2

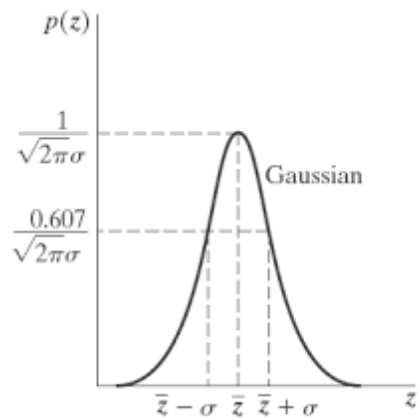

Figure2: Gaussian PDF[1]

**2.2 Poisson noise**

Poisson noise also known as photon noise is as a result of the statistical nature of electromagnetic waves such as x-rays, visible lights and gamma rays. These sources have random fluctuation of photons and resulting images have spatial and temporal randomness [5]. This noise obeys the Poisson distribution and is given as:

$$p(x) = \frac{\Lambda^x e^{-1}}{x!} \quad (5)$$

Where x = 0,1,2; e = Euler's constant ≈ 2.71828; $\Lambda$ = mean number of occurrences in the interval.

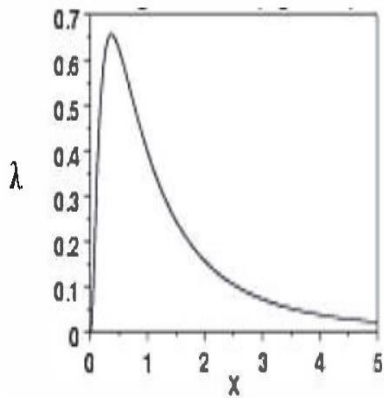

Figure3: Poisson distribution[1]

The mean and variance of the Poisson distribution are both equal to λ, which depends on the number of incident photons. So, the noise *b* depends on the noiseless image x.

## 2.3 Salt and Pepper noise

The salt and pepper noise is a type of impulse noise where the corrupt pixel takes either the maximum or minimum gray value, and appears as white and black pixels in the corrupt image [9]. An image having salt-and-pepper noise will have a few dark pixels in bright regions and a few bright pixels in dark regions. It can be caused by several factors such as dead pixels, analog-to-digital conversion error and bit transmission error. The probability density function is given as:

$$p(z) \begin{cases} p_a\ for\ z = a \\ p_b\ for\ z = b \\ 0\ otherwise \end{cases} (6)$$

Where *a* and *b* are usually *saturated* (maximum and minimum) values. If *b* > *a*, intensity *b* will appear as a light dot in the image. Conversely, *a* > *b*, level *a* will appear like a dark dot. The distribution is as shown:

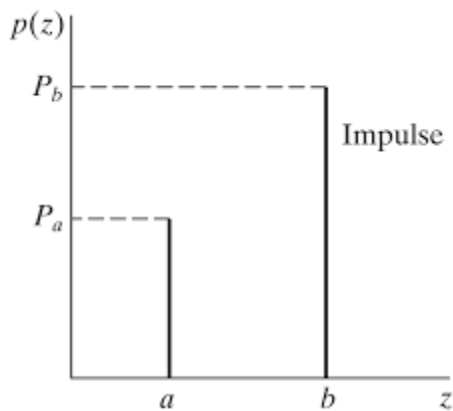

Figure 4: PDF of salt and pepper noise[1]

## 2.4 Speckle noise

A fundamental problem in optical and digital holography is the presence of speckle noise in the image reconstruction process. Speckle is a granular noise that inherently exists in an image and degrades its quality. Unlike Gaussian or Salt and Pepper noise, speckle noise is a multiplicative noise. It can be generated by multiplying random pixel values with different pixels of an image. Speckle noise model can be expressed as:

$$j = i + \mu * i \quad (7)$$

Where, j is the speckle noise distribution image, i is the input image and $\mu$ is the uniform noise image by mean o and variance v. The distribution and pixel representation of this noise is shown in figure 5.

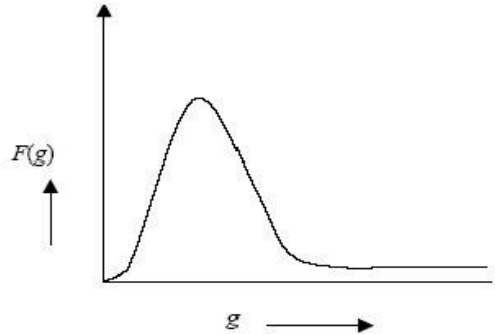

Figure 5: Speckle Noise[2]

**III. Different filtering techniques**

Image filtering is accepting or rejecting certain frequency components to either emphasize or remove other features. It could be high-pass or low pass filtering. A high pass filter tends to retain the high frequency information within the image while reducing the low frequency information causing image sharpening. Similarly, low pass filter allows low pass frequency information within the image. The net effect of low pass filter is blurring (smoothing) an image which causes reduction to image noise. This section discusses some spatial filters applied to images in this work. These include average filter, median filter, bilateral filter and Gaussian filter.

**3.1 Average (mean) filter**

Mean filter, or average filter is a windowed linear filter that smooths images or signals. The filter works as low-pass one. The basic idea behind this filter is that every element of the image takes an average across its neighborhood. This is achieved by simply replacing each pixel value in an image with the mean (`average') value of its neighbors, including itself. The effect of this is eliminating pixel values which are unrepresentative of their surroundings. This can be express by:

$$\hat{f}(x,y) = \frac{1}{mn}\sum_{(s,t)\in S_{xy}} g(s,t) \quad (8)$$

Where $s_{xy}$ represents the set of coordinates in a rectangular sub image window of size $m$ x $n$ centered at point (x,y). $g(x,y)$ is the corrupt image in the area defined by $s_{xy}$

**3.2 Median filter**

This a non-linear spatial filter that replaces the value of a pixel by the median of the intensity values in the neighbourhood of that pixel[1]. The median is calculated by first sorting all the pixel values from the window into numerical order, and then replacing the pixel being considered with the middle (median) pixel value. It preserves the edges while removing noise. Median filter performs well on both bipolar and unipolar impulse noise because of its appearance as white and black dots superimposed on an image. It is expressed as:

$$\hat{f}(x,y) = median_{(s,t)\in S_{xy}}\{g(s,t)\} \quad (9)$$

Where $s_{xy}$ represents the sub image area and $g(s,t)$ is the corrupt image.

**3.3 Gaussian filter**

Gaussian filter is a weighted average of the intensity of the adjacent positions. The weight decreases with the spatial distance to the center position. With Gaussian filter, image edges are blurred because pixels across discontinuities are averaged together. It is expressed as:

$$GC[I_p] = \sum_{q \in S} G_\sigma(||p - q||)I_q \quad (10)$$

Where $P$ is the center position. The weight for the pixel $q$ is defined by the Gaussian: $G_\sigma(||p - q||)$ and $\sigma$ is a parameter defining the neighbourhood size.

**3.4 Bilateral filter**

The bilateral filter is a nonlinear filter that performs spatial averaging without smoothing edges. It has shown to be an effective image denoising technique. Bilateral filter takes a weighted sum of the pixels in a local neighborhood. These weights depend on both the spatial distance and the intensity distance [16]. In this way, edges are preserved well while noise is averaged out. Mathematically, at a pixel location x, the output of a bilateral filter is calculated as follow:

$$BF[I_p] = \frac{1}{w_p} \sum_{q \in S} G_{\sigma_s}(||p - q||) G_{\sigma_r}(|I_p - I_q|)I_q \quad (11)$$

The parameters $\sigma_s$ and $\sigma_r$ will specify the amount of filtering for the image $I$. $W_p$ is a normalization factor.

**IV. Image Quality Evaluation Tools**

This section gives brief description of different evaluation tools applied to the restored images to determine the rate of similarity between the denoised image and the original image. There are several techniques and metrics available to be used for objective image quality assessment. These techniques are grouped into two categories:

1. Full-Reference (FR) approaches: The FR approaches focus on the assessment of the quality of a test image in comparison with a reference image. This reference image is considered as the perfect quality image that means the ground truth.

2. No-Reference (NR) approach: The NR metrics focus on the assessment of the quality of a test image only. No reference is made to the original image.

In this work, the full Reference approach is used as is considered a preferable quality assessment approach [12]. The full reference tools used in this work are the Mean Squared Error, Root Mean Square Error, Peak to Signal Noise Ratio, Structural Similarity Index Method and Universal Image Quality Index.

**4.1 The Mean Squared Error (MSE)**

This represents the cumulative squared error between the denoised image and the original image; the lower the value of MSE, the lower the error. MSE is given as

$$\frac{1}{MN}\sum_{M}^{i=0}\sum_{N}^{j=1}(f(i,j) - f'(i,j))^2 \quad (12)$$

In the above equation, $M$ and $N$ represent the number of rows and columns in the input images with index i and j respectively. *f(i,j)* represents the noisy image at location (i, j) and *f'(i,j)* represents the denoised image at location (i,j).

**4.2 Peak to signal noise ratio (PSNR)** PRSN defines the ratio between the maximum possible power of a signal and the power of corrupting noise that affects the quality of image [8]. PSNR represents a measure of the peak error and is given by:

$$PSNR = 10 log_{10}\left(\frac{(L-1)^2}{MSE}\right) \quad (13)$$

Where MSE is the mean squared error and L is the maximum possible intensity levels. The greater the PSNR value, the better the output image quality. That is, higher value of PSNR indicates that the reconstruction is of higher quality.

**4.3 The root mean square error (RMSE)**

RMSE computes the difference between two images. It evaluates the effectiveness of image reconstruction algorithm by computing the pixel-wise difference between the original image and the reconstructed image. The root mean square error is actually the square root of the Mean Square Error and is given as

$$RMSE = \sqrt{\frac{1}{MN}\sum_{m=0}^{M-1}\sum_{n=0}^{N-1}}$$

Where *M* and *N* represent the number of rows and columns in the input images with index *m* and *n* respectively. *f(m,n)* represents the original image at location *(m,n)* and *g(m,n)* represents the reconstructed image at location *(m,n)*. The smaller the RMSE value, the closer the reconstructed image to the original.

### 4.4 Structural Similarity Index Method (SSIM)

SSIM is used as a metric to measure the *similarity* between two given images. It can be expressed through three key features as:

$$SSIM(x,y) = [l(x,y)]^\alpha \cdot [c(x,y)]^\beta \cdot [s(x,y)]^\gamma \quad (15)$$

where, l is the luminance (*used to compare the brightness between two images*), c is the contrast (used to differ the ranges between the brightest and darkest region of two images) and s is the structure (*used to compare the local luminance pattern between two images to find the similarity and dissimilarity of the images*) and α, β and γ are positive constants [13]. This system calculates the *Structural Similarity Index* between 2 given images which is a value between -1 and +1. *A value of +1* indicates that the 2 given images are **very similar or the same** while a *value of -1* indicates the 2 given images are **very different.**

### 4.5 Universal image quality index (UQI)

UQI evaluates the quality of an image using loss of correlation, luminance distortion, and contrast distortion [10]. The measure is computed by comparing Riesz-transform features at key locations between the distorted image and its reference image. The range of this metric index *Q* is [-1, 1]. The value 1 indicates that the original and the reconstructed images are very similar or identical and -1 indicates otherwise. UQI is expressed as:

$$Q = \frac{\sigma_{fg}}{\sigma_f \sigma_g} \cdot \frac{2\overline{f}\overline{g}}{(\overline{f})^2 + (\overline{g})^2} \cdot \frac{2\sigma_f \sigma_g}{\sigma_f^2 + \sigma_g^2} \quad (16)$$

Where the first component of *Q* is the correlation coefficient and varies in the range [-1, 1], the second component with a value range of [0, 1] measures the luminance and the third component with value range of [0, 1] measures contrast similarity between the images.

### V. Results and Discussion:

The common types of noise encountered in image processing viz: Gaussian noise, Poisson noise, salt and pepper noise and speckle noise were applied to the same gray scale image using python. The effects are as shown in figures 6 and 7.

Table 1: Filters and image quality

| Filter | NOISE | RMSE | MSE | UQI | PSNR | SSIM |
|---|---|---|---|---|---|---|
| Average Filter | Gaussian | 0.05983119282508035 | 0.0035797711634871946 | 0.9808669295017859 | 24.46144677512086 | 0.808015030909535 |
| | Poisson | 0.05669687525280326 | 0.0032145356634319345 | 0.9857181518068635 | 24.928817516272446 | 0.835014581244546 |
| | Salt & Pepper | 0.06321413069209636 | 0.003996026319157439 | 0.9768561999637558 | 23.983716600288375 | 0.7869239814435646 |
| | Speckle | 0.05694718392281761 | 0.003242981756739217 | 0.9857399510178153 | 24.890554944259403 | 0.8326724279588097 |
| Gaussian Filter | Gaussian | 0.04970994355194228 | 0.002471078487937288 | 0.9849376136883428 | 26.071134600464397 | 0.849518724366986 |
| | Poisson | 0.042837590426082046 | 0.0018350591535127563 | 0.9918332329169578 | 27.363499316134675 | 0.9039183541044938 |
| | Salt &Pepper | 0.054883092593815054 | 0.0030121538526661277 | 0.9811270841568794 | 25.211228493188255 | 0.8197744424248894 |
| | Speckle | 0.04340574054611902 | 0.0018840583123570012 | 0.9918742096090828 | 27.249056597475136 | 0.9003161817622608 |
| Median Filter | Gaussian | 0.05766616335383758 | 0.00332538639595148 | 0.9827742605826306 | 24.781578842316158 | 0.7998444021220875 |
| | Poisson | 0.051134799433548046 | 0.0026147677131091855 | 0.9897493426560277 | 25.82566886297939 | 0.8500791660729071 |
| | Salt & pepper | 0.05018700683387147 | 0.0025187356549430614 | 0.9899662544244289 | 25.98817409962681 | 0.8627162124418586 |
| | Speckle | 0.05179463259341281 | 0.002682683965486621 | 0.9896900685720448 | 25.714304865265127 | 0.8456087835335383 |
| Bilateral Filter | Gaussian | 0.053652323018884646 | 0.0028785717653227396 | 0.9835065453631474 | 25.408229387309454 | 0.8274502296063884 |
| | Poisson | 0.04707494090616778 | 0.0022160500613191887 | 0.9901425763881235 | 26.544204329737944 | 0.88437529824273645 |
| | Salt & pepper | 0.05845377387313015 | 0.003416843680011033 | 0.9786251930577698 | 24.663748896418426 | 0.7977315920424216 |
| | Speckle | 0.04766747246155993 | 0.0022721879308735574 | 0.9901783186306675 | 26.435557513210423 | 0.8800595274921461 |

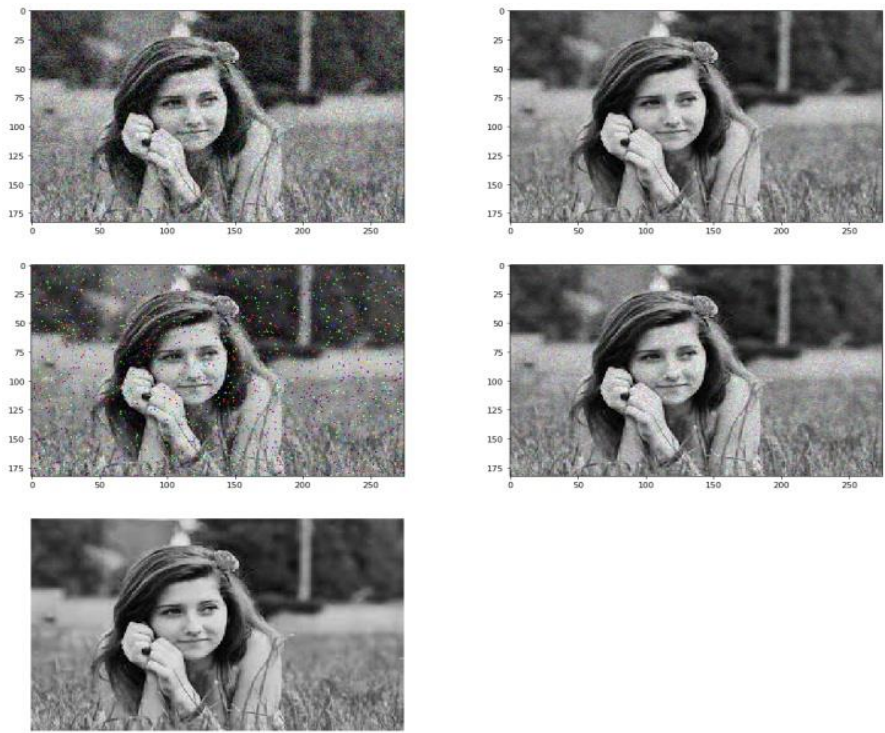

Figure 6: Application of noise models to same gray scale image. L-R: (A) Gaussian (B) Poisson (C) S&P (D) Speckle (E) Original

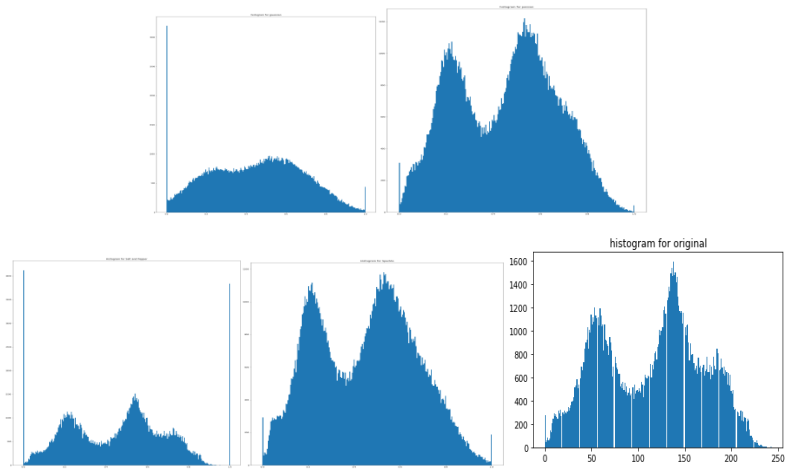

Figure7: Histogram of the noisy images. L-R: (A)Gaussian (B) Poisson (C) S&P (D) Speckle (E) Original

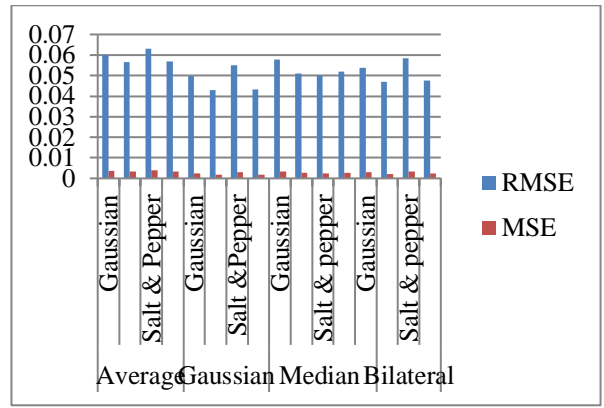

Figure 8: RMSE and MSE of various filters and Noise

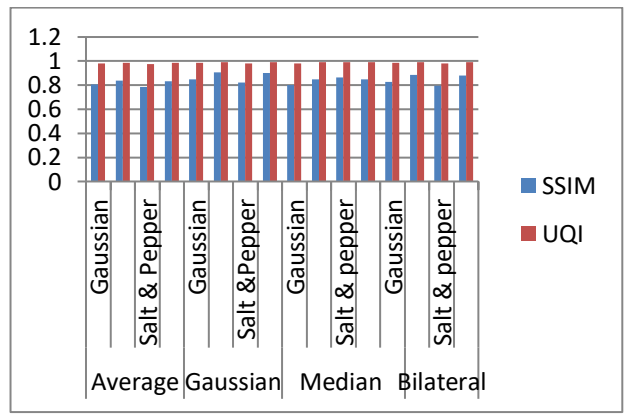

Figure 9: SSIM and UQI of various filters and Noise

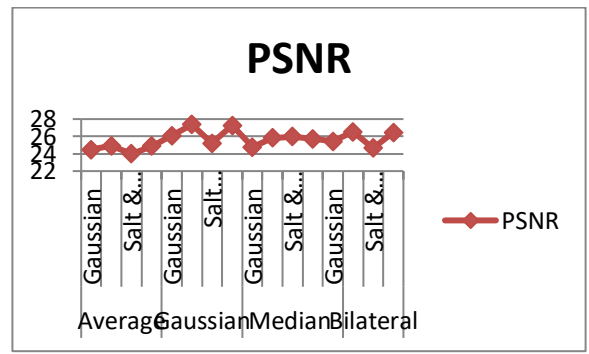

Figure 10: PSNR of various filters and Noise

We observed that for Gaussian noise, the whole image is affected in the same way by the noise (figures 6a and 7a). The histogram (fig 7a) shows that the noise tends to be evenly distributed causing the peak signals to be flattened when compared with the histogram of the original image (figure 7E). In case of the Poisson noise(figures 6b and 7b), the lighter parts are noisier than the dark parts. For impulse noise figures 6c and 7c, fewer pixels are modified and they are replaced by black and white pixels. Speckle noise(figures 6d and 7d) appeared as granular spots on the image.

Table 1 gives the result of applying different filtering techniques on different noise models. The values obtained shows how closer the reconstructed image is to the original image using: Peak to Signal Noise Ratio (PSNR), Root Mean Square Error (RMSE), Mean Square Error(MSE), Structural Similarity Index Method(SSIM) and Universal Image Quality Index (UQI). Lower RMSE and MSE value is an indication that the denoised image is closer to the original. Conversely, the higher the PSNR, the closer the reconstructed image to the original. For SSIM and UQI, its value ranges from -1 to +1. +1 shows that the reconstructed image is identical to the original and -1 otherwise. Thus, experiment shows that Gaussian filter is best suited for Gaussian noise, Poisson noise and speckle noise. It is less effective for salt and pepper noise. Next to Gaussian filter is the Bilateral filter which is not only effective in removing noise but also preserves edges of the image unlike Gaussian filter. Median filter happens to be best suited for salt and Pepper and it as well preserves sharp edges. Among the four filters, average filter tends to be the least effective on the sampled noise models.

## VI. Conclusion

Image quality can be affected adversely during acquisition or by the equipment used in capturing the image. In order not to affect the result of its intended use such as feature extraction, there is need to restore image to its original form. Different image filtering techniques were applied to noisy images in this work to determine the most effective filter for different types of noise.